\documentstyle[11pt]{article}
\def\be{\begin{equation}}
\def\ee{\end{equation}}
\def\bea{\begin{eqnarray}}
\def\eea{\end{eqnarray}}
\def\ba{\begin{array}}
\def\ea{\end{array}}
\def\al{\alpha}
\def\bt{\beta}
\def\ld{\lambda}
\def\sg{\sigma}
\def\gm{\gamma}

\begin{document}
\thispagestyle{empty}
\setcounter{page}0

~\vfill
\begin{center}
{\Large \bf Could HERA results have been predicted \\
from semileptonic meson decays}\\

\vspace{.5cm}
{\large M. V. Chizhov}\footnote{ On leave from 
Centre for Space Research and Technologies, Faculty of Physics,
University of Sofia, \\1164 Sofia, Bulgaria, E-mail: mih@phys.uni-sofia.bg}

\vspace{.5cm}

{\em
NORDITA, Blegdamsvej 17, DK-2100 Copenhagen \O, Denmark}
\end{center}
\vfill

\begin{abstract}
The anomalous value of the tensor form factor in the kaon decay 
$K^+ \to \pi^0 e^+ \nu$ and the destructive interference in pion decay
$\pi^- \to e^- \bar{\nu} \gamma$ can be simultaneously described by 
admixture of a tensor interaction in the standard $V$-$A$ one.
It is shown that the same tensor interaction can describe recent HERA
data at very large $Q^2$.
\end{abstract}

\vfill

\newpage

In the last years precision electroweak tests were performed at LEP, SLC
and Tevatron. The announcements about deviations from the Standard Model (SM)
have given rich possibility for theorists to speculate about the existence
of a new physics. Unfortunately, the anomalies seam to have disappeared 
after taking into account the systematic errors more precisely~\cite{ALEPH} 
and appropriately modifying the  structure functions~\cite{CTEQ4HJ}. 
Although some 
contradictions between the experimental data and the SM predictions exist,
they are not paid a great attention. Nevertheless, the recent HERA results
of H1~\cite{H1} and ZEUS~\cite{ZEUS} Collaborations have been met with 
a considerable interest. 

Now there are lots of theoretical works explaining the events excess at HERA
by some new physics. In our work we will present an explanation of the
deviation from the SM, observed in deep-inelastic scattering (DIS), by
introducing a new effective ({\it non contact}) interaction present in the
$t$-channel. Usually such interaction appears from the exchange of a new
intermediate boson with a large mass $M > M_{W/Z}$, which is almost
negligible for small momentum transfers. However, if the new boson has
approximately the same coupling constant with the fermions as 
$\gamma$, $W$ and $Z$ ones, then at high momentum transfer the strength of 
the new interaction will be comparable with the usual electroweak
interactions. In case of $e^+p \to e^+X$ DIS at large $Q^2$, there exists
a destructive interference between $\gamma$ and $Z$ exchange amplitudes
and the expected cross section is substantially reduced. Therefore, in
this case the additional contribution from the new interaction may be
more definitely observed than in $e^-p \to e^-X$ DIS.

Another characteristic feature of our approach is the suggestion about
the existence of new vector bosons, different from the gauge ones $W'$
and $Z'$, and inducing the new interaction. The new vector particles are
described by antisymmetric second-rank tensor field $T_{\mu\nu}$~\cite{Kemmer},
and not by the vector $A_{\mu}$. They correspond to the
representation of the Lorentz group (1,0) + (0,1) in contrast to the
widely used one (1/2,1/2) for the vector particles. As far as
both of these nonequivalent representations for the description of
vector particles exist in the Lorentz group on the same footing, we will
assume, that besides the gauge bosons, there exists another kind of vector
particles in Nature.

Due to the particular Lorentz indices structure, the new vector bosons
couple to the tensor currents $\bar{\Psi}_R\sigma_{\mu\nu}\Psi_L$,
$\bar{\Psi}_L\sigma_{\mu\nu}\Psi_R$, while the usual gauge bosons 
interact with the vector currents $\bar{\Psi}_L\gamma_{\mu}\Psi_L$,
$\bar{\Psi}_R\gamma_{\mu}\Psi_R$. This determines their main differences in
the interactions with fermions. Note, that while the gauge bosons
conserve the chiralities of the fermions, the new vector particles change
fermion chiralities. This feature allows to distinguish their presence
on the background of the standard electroweak interactions in
polarization experiments and by analyzing the angular distributions.

The helicity dependence of DIS cross section is parameterized by the
variable $y$. As far as at large $Q^2$ the contribution of the sea
quarks into the cross section is negligible, and scattering occurs mainly
on the valence quarks, the cross section of the charged-current (CC) 
process $e^+p \to \bar{\nu}X$ is suppressed by a factor $(1-y)^2$ in 
comparison with $e^-p \to \nu X$ due to chirality considerations. 
Therefore, the events excess, observed by H1 Collaboration~\footnote{
Unfortunately, ZEUS Collaboration has not announced its CC data.}
at large $y$, 
may indicate in favor of the presence of the new interactions leading 
to leptons and quarks spin flip.

Those interactions can be chosen in the form

\be
{\cal L}_T = \frac{g}{2\sqrt2}\left[
(\bar{\nu} \bar{e})_L \sg^{\al\bt} e_R + 
(\bar{u} \bar{d})_L \sg^{\al\bt} d_R \right]\left( 
\begin{array}{c}
T^+_{\al\bt}\\
T^0_{\al\bt}
\end{array}\right)
+\frac{g}{2\sqrt2}
(\bar{u} \bar{d})_L \sg^{\al\bt} u_R \left(  
\begin{array}{c}
U^0_{\al\bt}\\
U^-_{\al\bt}
\end{array}\right) + {\rm h.c.}
\label{Yukawa}
\ee

\noindent Here we have introduced two doublets of the new vector particles 
with opposite hypercharges, which are described by antisymmetric 
second-rank tensor fields.
These interactions conserve $SU(2)_L \times U(1)_Y$ symmetry of the 
electroweak interactions. For definiteness we choose the coupling 
constant of the new interactions equal to the gauge coupling constant
of the $SU(2)$ group. As usual, all particles are introduced
massless and the Lagrangian for the new particles reads

\be
{\cal L}_0 =
\frac{1}{4} (\partial_\ld T^{\mu\nu}_i)\partial^\ld T^i_{\mu\nu}
- (\partial_\mu T^{\mu\ld}_i)\partial^\nu T^i_{\nu\ld}
+
\frac{1}{4} (\partial_\ld U^{\mu\nu}_i)\partial^\ld U^i_{\mu\nu}
- (\partial_\mu U^{\mu\ld}_i)\partial^\nu U^i_{\nu\ld},
\label{free}
\ee

\noindent where $T_{\mu\nu}^i$ and $U_{\mu\nu}^i$ are real fields with
weak isospin index $i=1,2$. The quantization of these fields was 
described in \cite{queer}.
The chiral charged fields in (\ref{Yukawa})
are expressed through the latter ones as 
$T^+_{\mu\nu}=(T_{\mu\nu}^1+i \tilde{T}_{\mu\nu}^1)/\sqrt2$,
$T^0_{\mu\nu}=(T_{\mu\nu}^2+i \tilde{T}_{\mu\nu}^2)/\sqrt2$,
$U^0_{\mu\nu}=(U_{\mu\nu}^1+i \tilde{U}_{\mu\nu}^1)/\sqrt2$,
$U^-_{\mu\nu}=(U_{\mu\nu}^2+i \tilde{U}_{\mu\nu}^2)/\sqrt2$,
where $\tilde{T}_{\mu\nu}=\frac{1}{2}\epsilon_{\mu\nu\al\bt}T^{\al\bt}$,
$\tilde{U}_{\mu\nu}=\frac{1}{2}\epsilon_{\mu\nu\al\bt}U^{\al\bt}$ are
dual tensors.
The requirement for cancellation of anomalies can be fulfilled by introducing
also two doublets of the scalar Higgs fields $H_1=(H^0_1~H^-_1)$ and
$H_2=(H^+_2~H^0_2)$ with opposite hypercharges. After the spontaneous 
symmetry breaking, the new fields acquire masses, and mixing of the fields
may result. In this case the propagators for the charged fields 
$T^\pm_{\mu\nu}$ and $U^\pm_{\mu\nu}$ will have the form~\cite{MPL}

\bea
{\cal P}_{\mu\nu\al\bt}(q) &=& ~~~~~~~~~\left(
\ba{cc}
<T(T^-_{\mu\nu}T^+_{\al\bt})>_0 & <T(T^-_{\mu\nu}U^+_{\al\bt})>_0 \\
<T(U^-_{\mu\nu}T^+_{\al\bt})>_0 & <T(U^-_{\mu\nu}U^+_{\al\bt})>_0
\ea \right)
\nonumber \\
&=& \frac{4i}{\Delta(q)} \left(
\ba{cc}
(q^2-m^2)\Pi^-_{\mu\nu\al\bt}(q) & \mu^2 1^-_{\mu\nu\al\bt} \\
\mu^2 1^+_{\mu\nu\al\bt} & (q^2-M^2)\Pi^+_{\mu\nu\al\bt}(q)
\ea \right),
\label{propagator}
\eea

\noindent where $\Delta(q)=(q^2-m^2)(q^2-M^2)-\mu^4$, 
$1^\pm_{\mu\nu\al\bt}=(g_{\mu\al} g_{\nu\bt} - g_{\mu\bt} g_{\nu\al} 
\pm i \epsilon_{\mu\nu\al\bt})/4$, 
$\Pi^\pm(q)=1^\pm \Pi(q)=\Pi(q) 1^\mp$ and 
\be
\Pi_{\mu\nu\al\bt}=\frac{1}{2}(g_{\mu\al} g_{\nu\bt} - g_{\mu\bt} g_{\nu\al})
-{q_\mu q_\al g_{\nu\bt} - q_\mu q_\bt g_{\nu\al}
 -q_\nu q_\al g_{\mu\bt} + q_\nu q_\bt g_{\mu\al} \over q^2}
\label{Pi}
\ee

The masses of the new particles are defined through the mass parameters 
$\mu$, $m$ and $M$:
\bea
M^2_L &=& {M^2+m^2-\sqrt{(M^2-m^2)^2+4\mu^4} \over 2}
\nonumber \\
M^2_H &=& {M^2+m^2+\sqrt{(M^2-m^2)^2+4\mu^4} \over 2}.
\label{masses}
\eea

\noindent Analogous relations can be written also for the neutral
particles $T^0_{\mu\nu}$ and $U^0_{\mu\nu}$. Below we will give two 
arguments, which allow us to fix relations among $\mu$, $m$ and $M$,
and to express all quantities through single mass parameter only.

Using (\ref{Yukawa}) and (\ref{propagator}), we can write the effective
Lagrangian for the CC quark-lepton interaction

\be
{\cal L}^{CC}_{eff}=
-\frac{g^2}{\Delta(q)}\bar{e}_R\sg^{\al\ld}\nu_L~{q_\al q^\bt \over q^2}
~\bar{u}\sg_{\bt\ld}[(m^2-q^2)(1+\gm^5)+\mu^2(1-\gm^5)]d + {\rm h.c.}
\label{effCC}
\ee

\noindent Here we have used the identities

\bea
(1\pm\gm^5)\sg^{\al\bt}\otimes(1\pm\gm^5)\sg_{\al\bt} &\equiv&
{4 q_\al q^\bt \over q^2}
(1\pm\gm^5)\sg^{\al\ld}\otimes(1\pm\gm^5)\sg_{\bt\ld}
\nonumber \\
(1\pm\gm^5)\sg^{\al\bt}\otimes(1\mp\gm^5)\sg_{\al\bt} &\equiv& 0
\label{identities}
\eea

Due to kinematical reasons, the tensor interaction by itself does not
contribute into the semileptonic two-particle $\pi$-meson decay 
$\pi_{e2}$. However, it has been shown~\cite{Voloshin}, that
because of electromagnetic radiative corrections to the tensor
interaction (\ref{effCC}), the pseudotensor term 
$\bar{u}\sg_{\mu\nu}\gm^5 d$ generates an interaction between the 
lepton and pseudoscalar quark currents, to which the pion decay is very
sensitive~\cite{pseudoscalar}. Our model~\cite{MPL} allows to solve 
this problem in case the two massive parameters are assumed equal

\be
\mu^2=m^2.
\label{eq1}
\ee

\noindent Then, in case of $q^2 \ll m^2$, pseudotensor quark term
$\bar{u}\sg_{\mu\nu}\gm^5 d$ disappears from (\ref{effCC}), while tensor
term $\bar{u}\sg_{\mu\nu}d$ does not contribute into the decay of
the pseudoscalar pion, because of the parity conservation in the
electromagnetic interactions.

Another relation among the mass parameters can be derived analyzing the 
dependence of the new particles masses on the ratio $m^2/M^2$.
A remarkable characteristic of the eqs.~\ref{masses} at $\mu^2=m^2$
is the existence of a maximum value for the $M^2_L$, namely
for

\be
M^2=2.5~m^2
\label{eq2}. 
\ee

\noindent In the static limit this corresponds to the minimum of the energy
of interaction of the spinor particles, when they interact by exchanging 
the new vector particle with mass $M_L$. Then the masses
of the lighter $M_L=m/\sqrt2$ and the heavier $M_H=\sqrt3 m$ vector particles
occur related and can be expressed through the single mass parameter $m$.

In the low-energy limit $q^2 \ll m^2$ the effective interaction (\ref{effCC})
takes the form~\cite{MPL}

\be
{\cal L}^{CC}_{eff0}=-\frac{G_F}{\sqrt2} f_t~ \bar{e}\sg^{\al\ld}(1-\gm^5)\nu
~{q_\al q^\bt \over q^2}~\bar{u}\sg_{\bt\ld}d + {\rm h.c.},
\label{effCC0}
\ee

\noindent where the dimensionless constant

\be
f_t= \frac{\sqrt2}{G_F} \frac{g^2}{3 M^2_L}=\frac{8 M^2_W}{3 M^2_L}
\label{ft}
\ee

\noindent describes the strength of the new tensor interaction relative to 
the ordinary Fermi coupling. The analysis of the anomalies observed in 
radiative pion decay $\pi^- \to e^- \bar{\nu} \gm$~\cite{Bolotov} and 
three particles
kaon decay $K^+ \to \pi^0 e^+ \nu$~\cite{Akimenko} gives self-consistent
values of the constant $f_t=0.39\pm0.18$~\cite{MPL}. From eq.~\ref{ft}
the mass values of the new charged vector bosons can be estimated as
follows:

\be
M_L=230\pm56~{\rm GeV},~~~~~ M_H=563\pm137~{\rm GeV}.
\label{MLH}
\ee

 Now we already have everything necessary to determine the contribution
of the effective tensor interaction into the CC DIS cross section. 
Neglecting the sea quarks, in the lowest order of the parton model, the cross 
section can be simplified to

\be
{{\rm d}^2 \over {\rm d}x {\rm d}y} \left[
\ba{c}
\sg^{e^-p}_{CC} \\
\sg^{e^+p}_{CC}
\ea \right] = \frac{G_F^2 s}{2\pi} \left\{ [Y_+(y)\pm Y_-(y)] F_W(Q^2) +
f^2_t (1-\frac{y}{2})^2 F_T(Q^2) \right\} \left[
\ba{c}
u(x,Q^2) \\
d(x,Q^2)
\ea \right]
\label{CC}
\ee

\noindent where $Y_\pm(y)=1\pm(1-y)^2$ reflect the helicity dependence of 
electroweak interactions and 

\be
F_W(Q^2)=\left[{M_W^2 \over (M_W^2+Q^2)}\right]^2,~~~~~
F_T(Q^2)=(4 M_L^4 +2 M_L^2 Q^2 + \frac{1}{2} Q^4)
\left[{3 M_L^2 \over (M_L^2+Q^2)(6 M_L^2+Q^2)}\right]^2,
\label{CCFF}
\ee

\noindent are the form factors, which appear due to the finite masses of 
the intermediate bosons, and are normalized as $F_W(0)=F_T(0)=1$; 
$u(x,Q^2)$ and $d(x,Q^2)$ are the quark densities of the proton.

At large momentum transfer ($Q^2 > 15000$ GeV$^2$) with a total $e^+p$
integrated luminosity of 14.2 pb$^{-1}$, we predict $5.75\pm2.65$ 
events~\footnote{We have used GRV94 parameterization~\cite{GRV94} of 
quark distribution functions.} versus 4 events, observed by H1 Collaboration,
where $1.77\pm0.87$ are expected. Probably, the number of CC events, 
observed by ZEUS Collaboration with total luminosity of 20.1 pb$^{-1}$, 
should fall into the interval $8.05\pm3.75$.

In fig.~1 the plots of $x$ and $y$ distributions
of the CC cross sections, for $Q^2 > 15000$ GeV$^2$ and a fixed value 
$f_t=0.28$, are presented in comparison with the SM.
For this value of $f_t$ we obtain the total cross section of the
$e^+p \to \bar{\nu}X$ scattering $\sg_{tot}=0.28$ pb, 
which gives 4 events at the
total luminosity of 14.2 pb$^{-1}$. The considerable events excess for
large $Q^2$ is due to the chiral suppression of the processes 
$e^+p \to \bar{\nu}X$ in the SM. In the case of $e^-p \to \nu X$ scattering,
at $Q^2 > 15000$ GeV$^2$, we have obtained a total cross section equal
to $4.9\pm0.7$ pb in comparison with the expected in the SM~\footnote{ 
We assume that the error of 7-8\% for the SM predictions is mainly due to the 
uncertainties of the quark distribution of the proton.} $3.8\pm0.3$ pb.
Therefore, the contribution of the tensor interaction into the CC electron 
DIS will be negligible compared to the standard one.

In order to estimate the effect of the new tensor interactions into 
the neutral-current (NC) DIS, it is necessary to obtain analogous
formulae for the NC lepton-quark effective interactions. In contrast
to the CC processes, NC interactions at low energy are screened
by the electromagnetic ones. Moreover, as will be shown underneath,
the NC tensor interactions are parity-conserving and do not contribute
into the $P$ asymmetries.
Obviously, due to that, the admixture of the tensor forces
have not yet been observed in NC interactions.
We have not a direct experimental estimation for the NC tensor coupling
constant $f_t^{NC}$, in contrast to the CC case. However, as far as
both charged and neutral fields are components of one and the same 
$SU(2)$ multiplet, their masses should not differ noticeably,
just like the case of $W$ and $Z$ bosons. Moreover, at space like
momentum transfer $q^2 < 0$, in the $t$-channel, the form factors
have a smooth dependence on the intermediate boson masses. Therefore,
for our NC cross section estimations we will use the same mass values
(\ref{MLH}) and the same propagators (\ref{propagator}) as for charged
particles case.

The effective Lagrangian for the NC tensor interactions
between leptons and quarks has the form: 

\be
{\cal L}^{NC}_{eff}=-\frac{g^2}{2 \Delta(q)} \bar{e}\sg^{\al\bt} e~
[(q^2-m^2)\Pi_{\al\bt\mu\nu}(q)~\bar{d}\sg^{\mu\nu} d 
+ \mu^2\bar{u}\sg_{\al\bt} u].
\label{effNC}
\ee

\noindent Those interactions are parity conserving. Hence, they are 
not constrained from atomic parity violation measurements. 
Using the identities (\ref{identities}),
the Lagrangian (\ref{effNC}) can be rewritten in the chiral form

\bea
{\cal L}^{NC}_{eff}=&-&\frac{2 g^2}{\Delta(q)} \bar{e}_L\sg^{\al\ld} e_R~
{q_\al q^\bt \over q^2}[(m^2-q^2)\bar{d}_R\sg_{\bt\ld} d_L 
+ \mu^2\bar{u}_L\sg_{\bt\ld} u_R]
\nonumber \\
&-&\frac{2 g^2}{\Delta(q)} \bar{e}_R\sg^{\al\ld} e_L~
{q_\al q^\bt \over q^2}[(m^2-q^2)\bar{d}_L\sg_{\bt\ld} d_R 
+ \mu^2\bar{u}_R\sg_{\bt\ld} u_L].
\label{chiNC}
\eea

In the parton model the Born cross section for the NC scattering of 
leptons on the valence quarks are

\be
{{\rm d}^2 \over {\rm d}x {\rm d}y}\left[
\ba{c}
\sg^{e^-p}_{NC} \\
\sg^{e^+p}_{NC}
\ea \right]=\frac{2\pi \al^2 s}{Q^4}
\sum_{q=u,d}\left\{Y_+(y) C^q_2(Q^2) \pm Y_-(y) C^q_3(Q^2) +
(1-\frac{y}{2})^2 C^q_T(Q^2)\right\}~q(x,Q^2),
\label{NC}
\ee

\noindent where the form factors, $C^q_2$, $C^q_3$ and $C^q_T$ are
given by 

\bea
C^q_2(Q^2) &=& e^2_q-2e_q v_q v_e \chi_Z + (v_q^2+a_q^2)(v_e^2+a_e^2)\chi_Z^2
\nonumber \\
C^q_3(Q^2) &=& ~~~~-2e_q a_q a_e \chi_Z + (2v_q a_q)(2v_e a_e)\chi_Z^2
\nonumber \\
C^u_T(Q^2) &=& \left[\frac{\mu^2}{Q^2+2M_L^2}\right]^2 \chi_T^2,~~~~~
C^d_T(Q^2)=\chi_T^2
\label{coefs}
\eea

\noindent with

\be
\chi_Z={1 \over 4 \sin^2 \theta_W \cos^2 \theta_W} \frac{Q^2}{Q^2+M_Z^2},~~~
\chi_T={1 \over \sin^2 \theta_W} {2Q^2(Q^2+2M_L^2) 
\over (Q^2+M_L^2)(Q^2+6M_L^2)}.
\label{FFNC}
\ee

\noindent In eqs.~\ref{coefs} and \ref{FFNC}, $e_q$ is the quark charge
in units of $e$; $v_i$ and $a_i$ are the vector and axial couplings of the
fermions to $Z^0$; $\theta_W$ is the weak mixing angle. In the region 
of large $Q^2$ the contribution from the parity-violating term substantially 
reduces the $e^+p$ cross section. Therefore, for positron scattering
the events excess over the SM prediction is more significant.
The number of the observed and the predicted NC positron events, 
including tensor interaction, above $Q^2_{min}$ thresholds for the two 
experiments, are given in the table 1.

\begin{center}
\begin{tabular}{|c|r|c|} \hline
$Q^2_{min}$ [GeV$^2$] &	$N_{obs}$ &	$N_T$ \\ \hline
15000		& ~~24~~	& $30.7\pm 10.4$ \\
20000		& ~~10~~	& $12.9\pm 4.7$ \\
25000		& ~~6~~	& $5.7\pm 2.1$	\\
30000		& ~~ 4~~	& $2.6\pm 0.9$	\\
35000		& ~~2~~	& $1.1\pm 0.4$	\\ \hline
\end{tabular}
\end{center}

\noindent {\bf Table 1:} The number of the observed ($N_{obs}$) and 
the predicted ($N_T$) NC positron events, including tensor interaction, 
above $Q^2_{min}$ thresholds for the two experiments.\\ 

The differential cross sections with $Q^2 > 15000$ GeV$^2$ cut for NC DIS
are presented in fig.~2 for masses of the new particles

\be
M_L=248~{\rm GeV},~~~~~ M_H=607~{\rm GeV},
\label{MLHfix}
\ee

\noindent which correspond to $f_t=0.28$ (see eq.~\ref{ft}). 
For comparison the corresponding curves for the SM expectations are 
plotted also. The experimental data from the two experiments for positron 
scattering are shown in the figure as well.

The two experiments, with combined accumulated luminosity of 34.3 pb$^{-1}$
for positron scattering, have observed 24 events with $Q^2 > 15000$ GeV$^2$
against the SM expectation of $13.4\pm1.0$. For such a small number of 
events it is too early to speak about distributions and the total cross
section approach is more appropriate now. Our predictions for total 
cross sections at $Q^2 > 15000$ GeV$^2$ are $\sg^{e^-p}_{tot}=1.7\pm0.3$ 
pb and $\sg^{e^+p}_{tot}=0.9\pm0.3$ pb. 
For a total $e^+p$ integral luminosity of 34.3 pb$^{-1}$,
the last cross section corresponds to $30.7\pm10.4$
events. For our choice of the new particles masses (\ref{MLHfix}), this
corresponds to the total cross sections $\sg^{e^-p}_{tot}=1.5$ pb and 
$\sg^{e^+p}_{tot}=0.7$ pb (the last one being an exact fit to the observed
24 events).

So, our results on positron DIS are in agreement with anomalous
HERA data both for CC and NC channels. They reproduce the main
features of the HERA results, but it is too early to speak about
fitting various distributions due to the lack of sufficient
statistics.

Besides, note that the form of the new interaction was derived 
only using semileptonic meson decay data. And it is remarkable, that
this interaction works in self-consistent manner both at low and
high energies. The next step, of course, should be to analyze the 
eventual contribution from new interaction at hadron (Tevatron) and
lepton (LEP) colliders, which lies beyond the scope of the present
work. We would like to make just a short comment here. As far as
the masses of the new particles (\ref{MLHfix}) are large 
$M_H > M_L > m_{top}$,
it is very difficult to detect them in hadron collisions and a special
search for that is needed. However, up till now nobody has
searched for such kind of particles and interactions, neither at Tevatron 
nor at LEP. We hope that more precise analysis of the experimental data 
will allow to discover the effects of the new tensor interactions at both
Tevatron and LEP.

In conclusion, the author is glad to thank NORDITA, where this work 
has been fulfilled, for the hospitality
and especially Profs. J. Bijnens and  D. Diakonov for the understanding
and support. I would like to thank also Prof. A. Dolgov for the information 
about the most recent HERA results and discussions. I am also in 
debt to A. Sidorov for turning my attention to the fact that the 
tensor interactions discussed by me may have an effect in DIS.

\pagebreak[1]

\newpage
\vspace{2cm}
\begin{center}
{\bf FIGURE CAPTIONS}\\
\end{center}

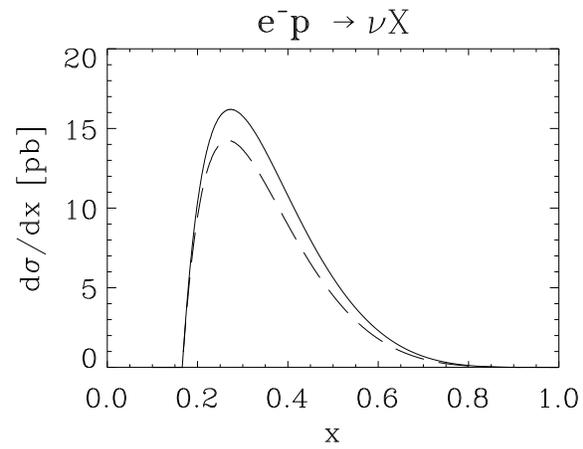
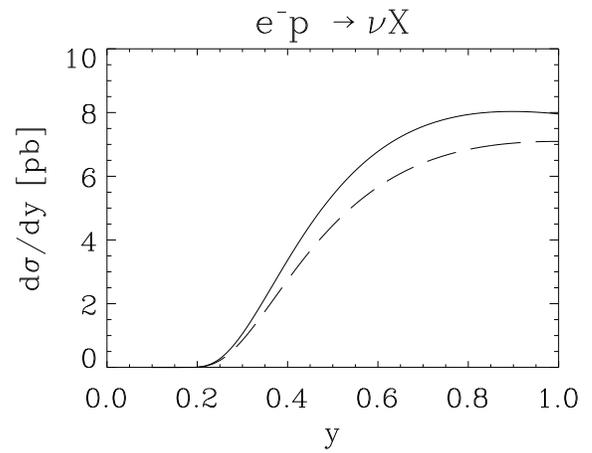
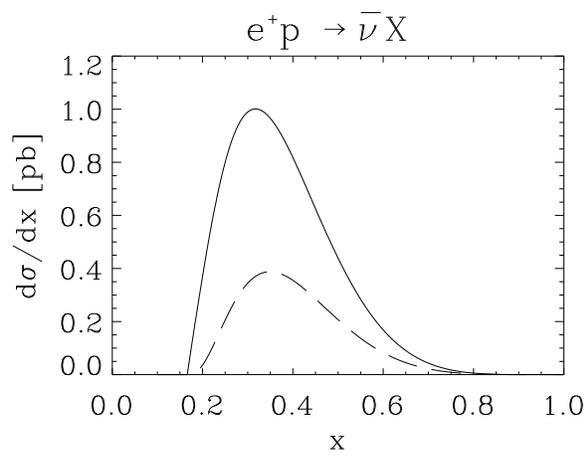
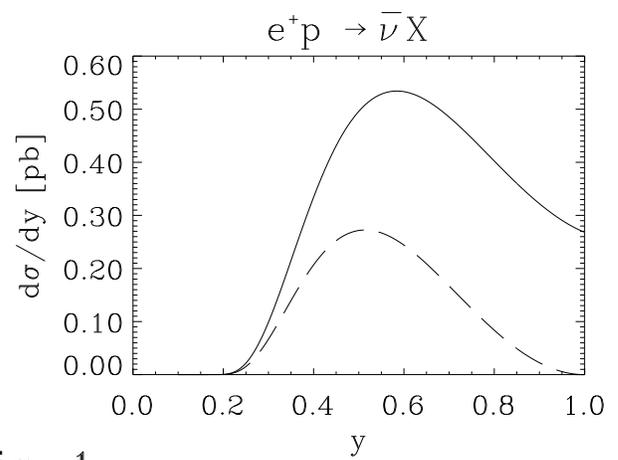
\noindent {\bf Figure 1:} 
The solid curves represent predicted CC differential cross 
sections for $Q^2 > 15000$ GeV$^2$, accounting for the tensor interaction 
with $f_t=0.28$. For a comparison the SM expectation curves are plotted 
with dashed lines.\\

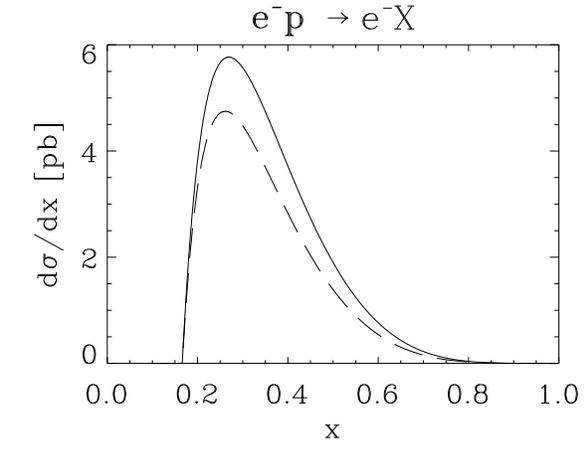
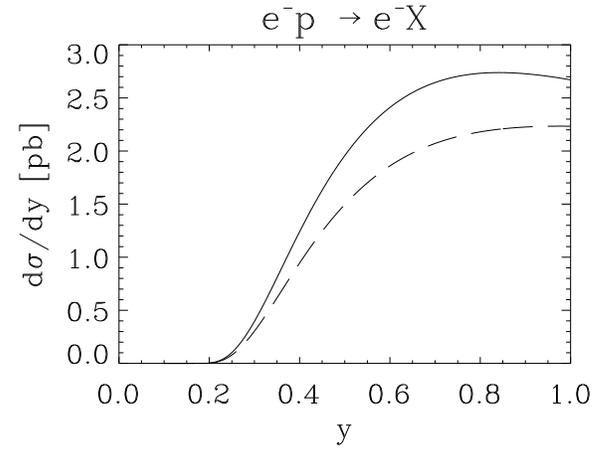
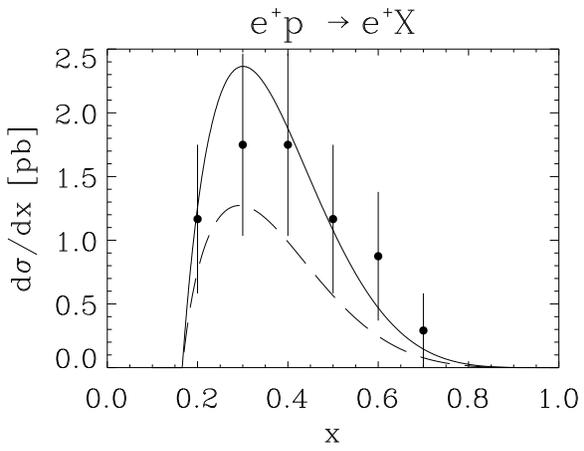
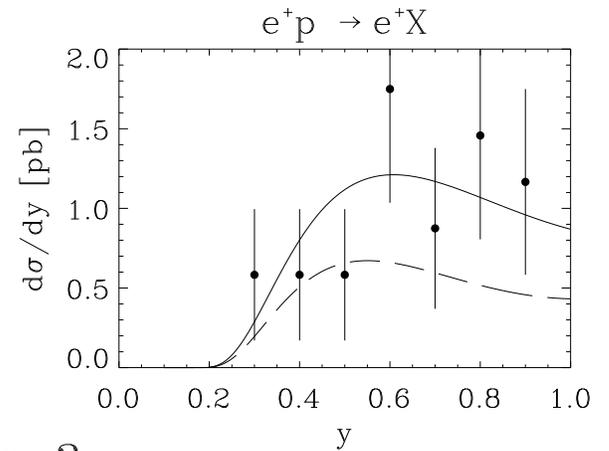
\noindent {\bf Figure 2:} 
The solid curves represent NC differential cross sections for 
$Q^2 > 15000$ GeV$^2$ and for masses of the new particles,
correspondingly $M_L=248$ GeV and $M_H=607$ GeV. For a comparison the 
SM expectation curves are plotted with dashed lines.
The experimental data from the two experiments for positron scattering
are shown with dots.

\end{document}